\documentstyle[twoside]{article}
\newcount\Mac  \Mac=0  
\newcommand{\ifMac}[2]{\ifnum\Mac=1 #1 \else #2 \fi}
\def\putps(#1,#2)(#3,#4)#5#6{\ifnum\Mac=1 \put(#1,#2){\special{picture #5}}
\else  \put(#3,#4){\includegraphics{#6}} \fi}
\def\Red  {}
\def\Black{}
\def\Blue {}
\newcommand{\U}{\,{\rm U(1)}}
\newcommand{\GeV}{\,{\rm GeV}}

\newcommand{\fig}[1]{~\ref{fig:#1}}
\newcommand{\lascia}[1]{}

\newcommand{\TeV}{\,{\rm TeV}}
\newcommand{\MeV}{\,{\rm MeV}}
\newcommand{\NP}{Nucl. Phys.}

\newcommand{\PRL}{Phys. Rev. Lett.}
\newcommand{\PL}{Phys. Lett.}
\newcommand{\PR}{Phys. Rev.}
\newcommand{\CL}{\,\hbox{\rm C.L.}}

\newcommand{\eq}[1]{~(\ref{eq:#1})}

\newcommand{\hc}{\hbox{h.c.}}
\newcommand{\md}[1]{\langle #1\rangle}

\newcommand{\sW}{s_{\rm W}}
\newcommand{\cW}{c_{\rm W}}

\def\Ord{{\cal O}}  \def\SU{{\rm SU}}
 
\def\circa#1{\,\raise.3ex\hbox{$#1$\kern-.75em\lower1ex\hbox{$\sim$}}\,}
\makeatletter
\def\art{\@ifnextchar[{\eart}{\oart}}
\def\eart[#1]#2#3#4#5#6{{\rm #2}, {\em #3 \rm #4} {\rm (#6) #5 ({\em #1})}}
\def\hepart[#1]#2{{\rm #2, \em#1}}
\newcommand{\oart}[5]{{\rm #1}, {\em #2 \rm #3} {\rm (#5) #4}}

\newcounter{alphaequation}[equation]
\def\thealphaequation{\theequation\hbox to
0.6em{\hfil\alph{alphaequation}\hfil}}
\def\eqnsystem#1{
\def\@eqnnum{{\rm (\thealphaequation)}}
\def\@@eqncr{\let\@tempa\relax \ifcase\@eqcnt \def\@tempa{& & &} \or
  \def\@tempa{& &}\or \def\@tempa{&}\fi\@tempa
  \if@eqnsw\@eqnnum\refstepcounter{alphaequation}\fi
\global\@eqnswtrue\global\@eqcnt=0\cr}
\refstepcounter{equation} \let\@currentlabel\theequation \def\@tempb{#1}
\ifx\@tempb\empty\else\label{#1}\fi
\refstepcounter{alphaequation}
\let\@currentlabel\thealphaequation
\global\@eqnswtrue\global\@eqcnt=0 \tabskip\@centering\let\\=\@eqncr
$$\halign to \displaywidth\bgroup \@eqnsel\hskip\@centering
$\displaystyle\tabskip\z@{##}$&\global\@eqcnt\@ne
\hskip2\arraycolsep\hfil${##}$\hfil& \global\@eqcnt\tw@\hskip2\arraycolsep
$\displaystyle\tabskip\z@{##}$\hfil
\tabskip\@centering&\llap{##}\tabskip\z@\cr}
\def\endeqnsystem{\@@eqncr\egroup$$\global\@ignoretrue} \makeatother

\begin{document}
\twocolumn[
\centerline{\bf hep-ph/9906266 \hfill  June 1999 \hfill   IFUP--TH/29--99}\vspace{1cm}
\centerline{\LARGE\bf\Red Bounds on Kaluza-Klein excitations of}\vspace{3mm}
\centerline{\LARGE\bf\Red   the SM vector bosons from electroweak tests}
\bigskip\bigskip\Black
\centerline{\large\bf Alessandro Strumia}\vspace{0.3cm}
\centerline{\em Dipartimento di fisica, Universit\`a di Pisa and INFN, sezione di Pisa,  I-56126 Pisa, Italia}
\bigskip\bigskip\Blue
\centerline{\large\bf Abstract}\begin{quote}\large\indent
Within a minimal extension of the SM in $4+1$ dimensions,
we study how Kaluza Klein excitations of the SM gauge bosons affect the electroweak precision observables.
Asymmetries in $Z$ decays provide the dominant bound on the compactification scale $M$ of the extra dimension.
If the higgs is so light that will be discovered at LEP2, we find the following $95\% \CL$ bounds:
$M>3.5\TeV$ (if the higgs lives in the extra dimension) and
$M>4.3\TeV$ (if the higgs is confined to our 4 dimensions).
In the second case, Kaluza Klein modes give `universal' corrections and a good fit of precision data can be obtained
with a heavier higgs (up to $500\GeV$) and with a smaller $M>3.4\TeV$.
\end{quote}
\Black\vspace{1cm}]
\noindent

\section{Introduction}
It has been recently realized that no known experimental constraint excludes extra spatial dimensions so large
that can be discovered at future experiments,
and that no known theoretical constraint excludes that this possibility be realized within string theory,
with a string scale in the TeV range~\cite{d}.
A scenario with such a low string scale can be motivated as a new possible solution of
the naturalness problem of having a  higgs much lighter than the Planck scale.

String excitations presumably generate a set of non renormalizable operators (NRO) suppressed by powers of the string scale.
Even assuming that only operators that conserve baryon number, lepton number, hadronic and leptonic flavour and CP are 
present, dimension 6  operators that affect the electroweak precision observables (EWPO) 
must be suppressed by a factor $1/\Lambda^2$ around $1/v^2N_Z^{1/2}$
(where $N_Z\sim10^7$ is the number of observed $Z$ decays and $v=175\GeV$).
A computation of all relevant operators~\cite{NRO} indeed shows that few of these operators
(and probably a generic set of  them)  must be suppressed by $\Lambda\approx10\TeV$.
We cannot however derive interesting implications from such bounds:
are they sufficiently strong to forbid observable Kaluza-Klein (KK) graviton effects~\cite{ColliderGravitonico} at LHC?
Is a higgs with $v=175\GeV$ natural if $\Lambda\circa{>}10\TeV$?
Similarly we cannot establish if observations about the 1987 supernova~\cite{supernova}, flavour violation, CP violation,
neutrino masses, proton decay, nucleosynthesis and cosmic baryon asymmetry are compatible with a so light string scale.
Unfortunately string theory is currently an example of a theory with no parameters that makes no calculable predictions.

In the following we will forget the possible but currently uncontrollable NRO of string origin
and we will study the bounds set by EWPO on the scale of
possible new extra dimensions where SM gauge bosons propagate.
This kind of extra dimensions are interesting because, if larger than the string scale,
modify  the string predictions for the gauge couplings 
in a way that qualitatively resembles the observed values~\cite{DDG}.
Beyond affecting the parameters of the SM,  KK excitations of the gauge bosons also
give minimal computable corrections to EWPO~\cite{NY,MP,Marciano}.

In section 2 we briefly recall a concrete minimal extension of the SM to 5 dimensions~\cite{M5SM} (`M5SM') and we
write the effective Lagrangian below the compactification scale $M$
in terms of the complete set of non renormalizable operators that affect EWPO used in~\cite{NRO}
and listed in the appendix.
Since we do not know if the higgs field should be confined to our 4 dimensions
or can propagate in the extra dimensions, the model contains two higgs doublets
with the two different behaviors.
Only one unknown parameter is necessary to take into account the
resulting uncertainty in low energy effects~\cite{MP}.

\begin{figure*}[t]
\begin{center}
\begin{picture}(17.7,5)
\putps(0,0)(0,0){h4d}{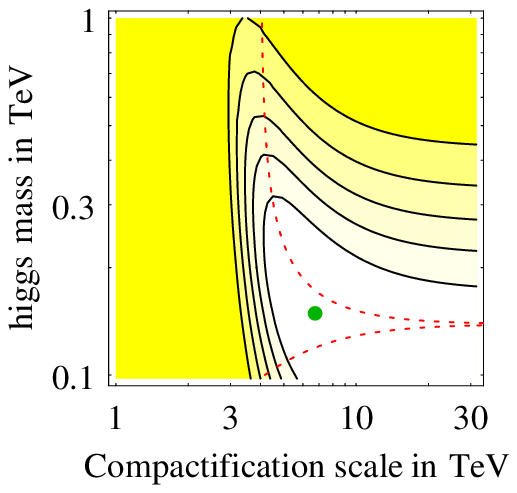}
\putps(6,0)(6,0){h5d}{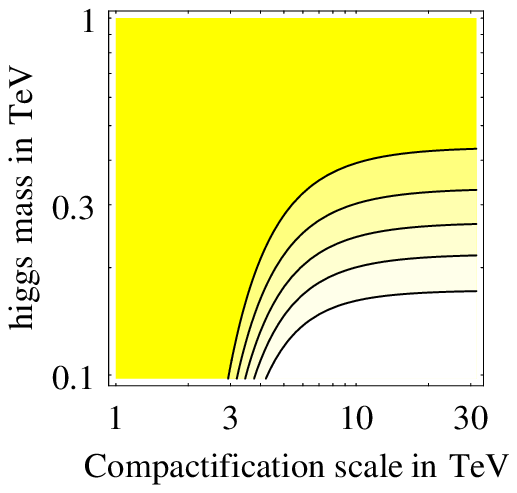}
\putps(12,0)(12,0){h45d}{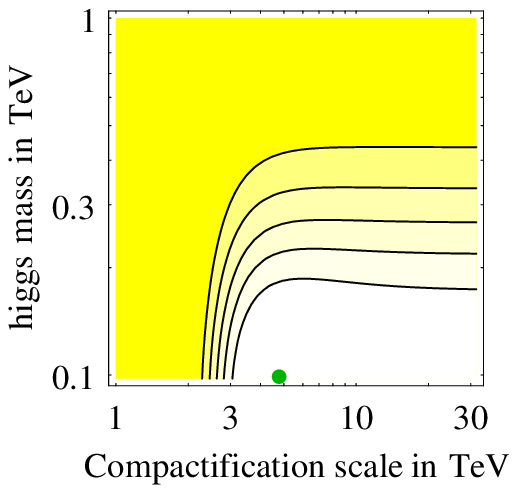}\Red
\put(3.2,5.4){\makebox(0,0)[c]{Higgs in 4d ($\sin \beta=1$)}}
\put(9.2,5.4){\makebox(0,0)[c]{Higgs in 5d ($\sin \beta=0$)}}
\put(15.2,5.4){\makebox(0,0)[c]{$\sin \beta=0.64$}}\Black
\end{picture}
\caption[SP]{\em Bounds on the compactification scale $M$ and on the higgs mass
in three extensions of the SM to five dimensions:
(a) with a higgs doublet confined to 4 dimensions;
(b) with a higgs doublet that can propagate to the extra dimension;
(c) a combination of (a) and (b).
The level curves correspond to $\Delta \chi^2=\{1,2.3,3.8,6,9.2\}$.
\label{fig:vincoli}}\end{center}\end{figure*}

In section~3 we derive bounds on $M$ from a global fit of the most recent data
about electroweak precision observables~\cite{mh<250,diLella}, listed in table~1.
The list of observables includes the Fermi constant measured in $\mu$ decay and
the $Z$ mass (known with great precision),
the $W$ and top masses, the various $Z$ widths, 
the various asymmetries in $Z$ decays grouped into an `effective $\sW$'
and the values of the electromagnetic and strong gauge couplings.
The list does not include atomic parity violation (APV), the neutrino-nucleon
cross sections and tests of quark-lepton universality.
Their inclusion would not shift our final results (the best-fit regions in the $m_h$ and $M$ plane shown in fig.\fig{vincoli})
in a significant way;
however,
since there is now some discrepancy between the measured value of APV~\cite{Csexp,QWexp}
and the SM prediction,
the inclusion of APV would  strongly deteriorate the quality
of the SM and M5SM fits~\cite{QWfit}\footnote{Recent measurements about the
Cesium atomic structure~\cite{Csexp} correct previous data and
allow to reduce the  atomic structure uncertainties, that still remain the largest uncertainty in the SM prediction for APV in Cesium.
The measured value of APV~\cite{QWexp},
$Q_W=-72.06\pm0.28_{\rm exp}\pm0.34_{\rm th}$, is now significantly smaller in modulus
(2.5~`standard deviations', if one adds experimental and theoretical errors in quadrature) than the SM prediction.
Not including APV in the fit we  derive more safe bounds on the mass of KK modes (that increase the amount of APV).
Its inclusion would strongly increase the minimal value of the $\chi^2$, both in the SM and in the M5SM,
and would allow to derive strong bounds on these models~\cite{QWfit} based on the `goodness of the fit'.
It would instead not shift in a significant way the 
`confidence intervals' on $m_h$ and $M$ that we study in this paper:
the $0.6\%$ error on APV is still significantly larger than the $\sim0.1\%$ error
on various cleaner electroweak observables listed in table~1.
For example, in a pure SM fit, the inclusion of APV  reduces by only $8\GeV$ (i.e. by $\sim 7\%$) the best fit value of $m_h$
while increases by $\sim250\%$ the value of $(\min\chi^2)/\hbox{d.o.f.}$.
`Confidence intervals' are commonly employed to report experimental data, in place of intervals based on `goodness of the fit',
due to their stability with respect to rare statistical fluctuations and/or underestimated systematic errors.}.

\begin{table}[t]
$$\begin{array}{rll}
M_Z &=& 91.187\GeV\\
G_\mu &=&1.1664~10^{-5}~\GeV^{-2}\\
\Gamma_Z &=&(2.4939\pm0.0024)\GeV \\
R_h  &=& 20.765\pm0.026 \\
R_b  &=& 0.21680\pm0.00073 \\
\sigma_h &=& (41.491\pm0.058)\hbox{nb} \\
s^2_{\rm eff}  &=& 0.23157\pm0.00018 \\
M_W  &=&(80.394\pm0.042)\GeV \\
m_t  &=& (174.3\pm5.1)\GeV\\
\alpha_3(M_Z)  &=& 0.119\pm0.004 \\
\alpha_{\rm em}^{-1}(M_Z)  &=& 128.92\pm0.036 \\
\end{array}$$
\caption{\em Electroweak precision observables.}
\end{table}

\section{The model}
In this section we briefly recall a concrete minimal extension of the SM to 5 dimensions~\cite{M5SM} 
and we compute how the KK excitations of the SM gauge bosons affect the EWPO.
The model contains one extra dimension compactified on ${\cal S}_1/{\cal Z}_2$
where the SM gauge fields can propagate
(the circle ${\cal S}_1$ has radius $R=1/M$; the ${\cal Z}_2$ symmetry ensures
that the massless spectrum only contains the SM fields).
The SM fermions are instead confined to 4 dimensions.
The higgs doublet could follow both possibilities.
Since we do not know which possibility (if any) is the physical one,
the model has two higgs doublets $H_4$ and $H_5$:  $H_4$ is confined to our 4 dimensions
while $H_5$ can propagate into the extra dimension.
Both higgs doublets could contribute to EWSB.
Their effects can be parameterized in terms of an angle $\beta$~\cite{MP}
$$\md{H_4}=(0,v\sin\beta),\qquad\md{H_5}=(0,v\cos \beta)$$
where $v=175\GeV$ and $\beta$ has nothing to do with the $\beta$ used in supersymmetric models.

At tree level the KK excitations $A^n_\mu$ of the SM gauge fields, with mass $M_n=n M$ ($n=1,\ldots,\infty$), couple to
SM particles with the Lagrangian interaction $\sqrt{2} A_\mu^n J_\mu$
where $J_\mu$ are the contributions from four dimensional fields to the usual gauge currents.
Conservation of momentum in the extra dimension forbids the five-dimensional fields to appear in the currents
(see~\cite{NY,MP} for more details).
All our analysis of precision data could be rephrased in terms of excited gauge bosons $A^*_\mu$ with couplings $g^*=\sqrt{2}g$.
Since KK modes are currently  more popular than composite particles,
we will perform our analysis with the normalization factors appropriate for KK modes.

\smallskip

The fact that KK modes couple to observed particles in a way similar to the SM vector bosons
has been used to compute the KK corrections to various EWPO~\cite{NY,MP,Marciano}.
Here we follow a less direct strategy because we prefer to use the results in~\cite{NRO}, where
the effects of a complete set of 10 non renormalizable operators (recalled in the appendix) on EWPO have been listed.

\smallskip

Thus we need to write the effective Lagrangian for the SM fields
obtained integrating out the KK excitations.
The first KK level gives
\begin{equation}\label{eq:JJ}
{\cal L}_1=-\frac{1}{M^2}( J^a_\mu J^a_\mu +  J^B_\mu J^B_\mu + J^G_\mu J^G_\mu)
\end{equation}
In this model the $n$-th KK mode gives (at tree level) ${\cal L}_n={\cal L}_1/n^2$,
so that summing over $n$ one obtains ${\cal L}_{\rm eff}=\frac{\pi^2}{6}{\cal L}_1$.
With one extra spatial dimension the first few terms dominate.
With more than one extra dimension the sum over KK excitations is divergent.
In both cases the string scale cannot be much larger than the compactification scale
and will cutoff the sum at $n\circa{<} M_{\rm string}/M$.
In general we do not know the numerical factor that relates
${\cal L}_{\rm eff}$ to ${\cal L}_1$.
In order to avoid a normalization of $M$ different from
previous analysis~\cite{MP}, we keep the factor $\pi^2/6$.
This model dependent normalization factor is not much relevant (but not completely irrelevant) when
comparing  LEP1 bounds with capabilities of  LHC.

The gluonic current does not affect electroweak precision observables so that we neglect it in the following.
The currents coupled to the KK modes of the $\SU(2)_L$ and $\U_Y$ gauge vector bosons are
\begin{eqnarray*}
J^a_\mu&=&\frac{g_2}{2}\bigg[\sum_{L,Q}(\bar{F}\gamma_\mu \tau^a F)+(i~H_4^\dagger  \tau^a D_\mu H_4+\hc)\bigg]\\
J^B_\mu&=&g_Y\bigg[\sum (Y_F \bar{F}\gamma_\mu F)+Y_H(i~H_4^\dagger  D_\mu H_4+\hc)\bigg]
\end{eqnarray*}
The sum in $J^a$ runs over the fermionic doublets $F=L,Q$, while the sum in $J^B$ runs over all the SM fermions.
In the standard notation that we employ the hypercharges $Y_F$ are
$$\{Y_L,Y_Q,Y_E,Y_U,Y_D\}=\{-\frac{1}{2},\frac{1}{6},-1,\frac{2}{3},-\frac{1}{3}\}.$$
The effective Lagrangian can thus be written in terms of the operators in the appendix as
\begin{eqnarray}
 {\cal L}_1 &=& \nonumber
-\frac{g_2^2}{2M^2 }\bigg[-\Ord_{LL}+\sin^2 \beta(\Ord'_{HL}+\Ord'_{HQ})\bigg]+ \\
&&-\frac{g_Y^2}{M^2}\sin^2 \beta \bigg[\sum Y_F \Ord_{HF}+ \Ord_{H}\sin^2 \beta\bigg]\label{eq:LO}
\end{eqnarray}
up to operators that do not affect the electroweak precision observables that we consider.
Using the results in~\cite{NRO} it is straightforward to compute the corrections from this set of NRO
to the various EWPO.

\medskip

Two limiting cases are of particular interest.
If $\sin\beta=0$ (EWSB entirely due to a five-dimensional Higgs $H_5$)
the relevant effective Lagrangian contains a single operator
$${\cal L}_1=\frac{g_2^2}{2M^2}\Ord_{LL}$$
again up to terms that do not affect EWPO
so that the bounds on it are the same as those in~\cite{NRO}.
The limiting case $\sin\beta=0$ is however problematic since it seems unlikely that 
the Yukawa couplings of a 5 dimensional Higgs to the top quark,
suppressed by a factor $\sim R^{1/2}$, can generate the top mass.

More interesting is the opposite limit, $\sin\beta=1$
(EWSB entirely due to a four-dimensional Higgs $H_4$; $H_5$ is either absent or irrelevant).
In this case the effective Lagrangian can be rewritten in the simple form
\begin{equation}\label{eq:WWeBB}
 {\cal L}_1  = -\frac{1}{M^2}(\Ord_{WW}+\Ord_{BB})
\end{equation}
once again up to terms that do not affect EWPO\footnote{We explicitly demonstrate this fact in the abelian case.
Applying the Bianchi identities $\partial_\alpha B_{\mu\nu}=-\partial_\mu B_{\nu\alpha}-\partial_\nu B_{\alpha\mu}$
to one of the two factors in $\Ord_{BB}=\frac{1}{2}(\partial_\alpha B_{\mu\nu})(\partial_\alpha B_{\mu\nu})$
and integrating by parts,
one obtains products of the combinations $\partial_\alpha B_{\alpha\nu}$ that
appear in the classical equation of motion of the $B_{\alpha\nu}$ gauge field.
In this way one obtains
 $\int d^4x~\Ord_{BB}=\int d^4x~J_\mu^B J^B_\mu$
up to surface terms.}
Beyond proving a check of the computation, eq.\eq{WWeBB}
will be useful for interpreting the bounds that we will find in the limiting case $\sin\beta=1$.
An interesting analysis of these two operators can be found in~\cite{GW}.

\medskip

It would be simple to extend the analysis to the more general models considered in~\cite{Carone}:
for example if also the leptons can propagate in the extra dimensions one
should omit all operators involving leptons from the effective  Lagrangian\eq{LO};
if instead the $\SU(2)_L$ gauge bosons are confined to our 4 dimensions one should omit their $J_\mu^aJ_\mu^a$ contribution.

\begin{figure}[t]
\begin{center}
\begin{picture}(17.7,5)
\putps(-0.5,0)(-0.5,0){sb}{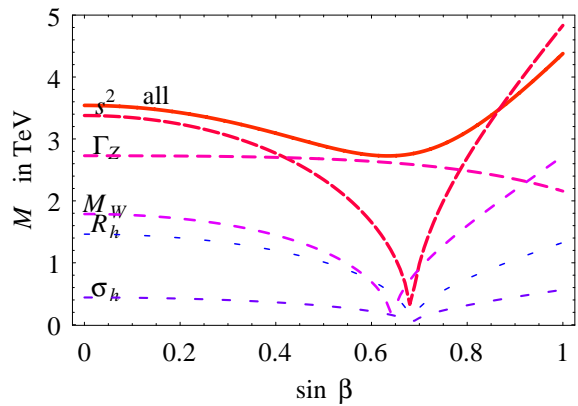}
\end{picture}
\caption[SP]{\em $95\%\CL$ lower bounds on the compactification scale $M$
from electroweak data for $m_h=100\GeV$
as function of $\sin\beta$
($\sin\beta=1$ corresponds to a higgs confined to 4 dimensions).
The continuous line shows the bound from the global fit,
while dashed lines show the bounds from single measurements.
\label{fig:sb}}\end{center}\end{figure}

\section{Results}
We begin our analysis making the simplifying assumption that
the higgs is so light that will be observed at LEP2 or Tevatron in the next years.
For fixed $m_h\approx100\GeV$, we can compute a $\chi^2(M,\sin\beta)$ by minimizing the full $\chi^2$
with respect to $m_t$, $\alpha_{\rm strong}$ and $\alpha_{\rm em}$
for each value of $M$ and $\sin\beta$.

In fig.\fig{sb} we show the resulting $2\sigma\approx 95\%\CL$ bound (i.e.\ $\Delta \chi^2=3.85$)
on the KK mass $M$ as function of $\sin\beta$ (continuous line).
The dashed lines show the bounds from the single experimental data that we have fitted (omitting the less relevant ones).
The apparent difference with respect to an analogous fig.\ in~\cite{MP} is only due to a different choice
of a minimal set of experimental data (with our choice there are no significant correlations between the data).
The strongest bound comes from the `effective $\sW$' extracted at LEP and SLD from various asymmetries in $Z$ decays.
Due to the accidentally small SM value of the  vector
coupling of leptons to the $Z$, these asymmetries are particularly sensible to new physics.
More precisely KK modes affect the correlation between this `effective $\sW^2$'  and the $\sW^2$ derived from
$\alpha_{\rm em}$,
the $Z$ mass, $M_Z$, and the Fermi constant for $\mu$ decay, $G_\mu$,
as $\sW^2\cW^2=\pi \alpha_{\rm em}/\sqrt{2}G_\mu M_Z^2$.

From fig.\fig{sb} we see that the $95\%\CL$ bound on the compactification scale $M=1/R$
is $M>3.5\TeV$ in presence of one higgs doublet that propagates in the extra dimension,
and $M>4.3\TeV$ in the simplest case of EWSB entirely due to a higgs confined to four dimensions.
Asymmetries in $Z$ decays give the dominant bound, except when $\sin^2\beta$ is close to $\sqrt{2}-1$,
where some cancellations take place.
In this case the measurement of the $Z$ width provides the weaker bound $M>2.7\TeV$.

\medskip

It is interesting to perform a more complicated analysis and study the EWPO bound on $M$ in association
with the EWPO bound on the higgs mass.
Since EWPO imply a light higgs only in absence of new physics,
both the upper bound on $m_h$ and the lower bound on $M$ could be somewhat relaxed.
Few cases where this happens can be found in~\cite{Casalbuoni,HK,NRO}.
This happens also in the case that we are studying if $\sin\beta$ is large enough
(i.e.\ in the simplest scenario with only a 4-dimensional higgs).
In this case KK corrections decrease the predicted value of the `effective $\sW^2$' that gives the strongest bound on $M$,
while a heavy higgs increases it.
The measurement of the effective $\sW^2$ thus allows to  have light KK modes and a heavy higgs,
with an appropriate cancellation of their effects.
Since other significant observables, like the $Z$ width,
do not allow such cancellation, only a limited weakening of the limits on $M$ and $m_h$ is possible.
The contour plots of the $\Delta \chi^2(M,m_h)$ in fig.s\fig{vincoli} show
the allowed regions in the $(M,m_h)$ plane  for different values of $\sin\beta$.
Fig.\fig{vincoli}a refers to a higgs confined to four dimensions;
while fig.\fig{vincoli}b is valid in the opposite (problematic) limit where only a 5 dimensional higgs exists.
No cancellations between heavy higgs and KK effects happen in this second case.
Finally fig.\fig{vincoli}c refers to the case $\sin^2\beta=\sqrt{2}-1=0.64$ that has the weakest bound
in fig.\fig{sb} due to some cancellations between KK effects.

The $\Delta \chi^2(M,m_h)$ plotted in fig.s\fig{vincoli} is precisely defined as follows.
For a given value of $\sin\beta$, and for each value of $M$ and $m_h$ we compute a $\chi^2(M,m_h)$ minimizing the full $\chi^2$ with respect to $m_t$, $\alpha_{\rm strong}$
and $\alpha_{\rm em}$.
$\Delta \chi^2$ is defined as $\chi^2(M,m_h)$ minus its minimum value, whose location is marked in fig.s\fig{vincoli} with a disk.
The chosen contour levels correspond to the conventional $68\%$, $95\%$ and $99\%$ `confidence levels'
on $m_h$, on $M$ or on the couple $(m_h,M)$.\footnote{These results can be easily translated into the results of a Bayesian analysis.
The contour levels correspond  to values of $-2\ln p$, where $p$ is the  Bayesian probability density (normalized to be one at the best fit point)
in  $\ln m_h$ and $1/M^2$ obtained assuming a flat prior distribution in the same variables.}

Fig.\fig{vincoli}a also contains a dashed line.
It has been plotted to illustrate how natural is having a higgs lighter than the compactification scale.
At the right of the dashed line, the quadratically divergent one loop correction to the squared higgs mass,
computed in the SM~\cite{dmhSM} and cutoffed at $M$, is more than ten times larger than the squared higgs mass itself
(so that a fine tuning $\circa{>}10$ is required).
We see that from the point of view of this naturalness problem,
a heavy higgs is not more natural than a light one, due to its larger self-coupling
(the particular behavior of the dashed line for $m_h$ around $130\GeV$ is due to a cancellation between the various SM loop effects).
Supersymmetry could still be necessary to justify the lightness of the higgs.
No change in our analysis is necessary in the supersymmetric case,
since experimental bounds on sparticles~\cite{diLella,pdg} guarantee that the EWPOs more crucial for our analysis
are not affected by significant supersymmetric loop effects~\cite{ABC}.
Maybe supersymmetry could force the higgs to be light.
This is however not guaranteed in presence of non renormalizable operators:
the superpotential can contain
a quartic term $W_4= (H_{\rm u} H_{\rm d})^2/2\Lambda$,
and the Lagrangian can contain the corresponding `$A$-term' $A W^{(4)}$.
Here $H_{\rm u}$ and $H_{\rm d}$ are the two higgs superfields required by supersymmetry.
If $\Lambda\sim$ (few TeV) and the soft terms $\mu,A$ are larger than $v$, the light higgs mass $m_h$ could receive significant corrections.
In this case it is easy to compute $m_h$ analytically using the results (and the notations) of~\cite{2h},
where the general two-higgs-doublet potential is studied.
The NRO terms induce extra contributions to the quartic couplings
$$\delta \lambda_6=\delta\lambda_7=\frac{\mu^*}{\Lambda},\qquad
\delta \lambda_5=-\frac{A}{\lambda}.$$
The light higgs is no longer constrained to be smaller than $M_Z|\cos2\beta|$ at tree level as in the renormalizable case:
$|\mu/\Lambda|\circa{>}0.1$  is sufficient for having $m_h\circa{>}100\GeV$ even at $\tan\beta=1$.

\section{Conclusion}
Within a minimal extension of the SM to 5 dimensions we have studied the bounds
on the size of the compactified extra dimension,
due to minimal corrections to EWPO mediated by KK excitations of the SM vector bosons.
If the higgs is so light that will be discovered at LEP2
(a possibility suggested by EWPO themselves and by supersymmetric models),
we find the following $95\%\CL$ bounds  on the radius $R=1/M$ of
the extra dimension where gauge fields can propagate:
$M>3.5\TeV$ (if the higgs lives in the extra dimension)
$M>4.3\TeV$ (if the higgs is confined to our four dimensions).

In the second case the bound can be a bit relaxed because
KK corrections allow a good fit of all precision data even in presence
of a heavier higgs, up to $\sim500\GeV$.
Accidental cancellations that compensate in many precision observables the loop corrections
of a heavy higgs with the KK corrections
are not very unlikely in a case, like this one,
where both corrections affect the
precision observables in an `universal' way.
This means that all the effects can be confined to the propagators of the gauge bosons (as in eq.\eq{WWeBB}) so that
all the experimental data are affected only through few parameters
(usually called $S,T,U$~\cite{STU} or $\epsilon_1,\epsilon_2,\epsilon_3$~\cite{BCF}).
No compensation happens in $\epsilon_1$  (that is reduced both by heavy higgs and KK corrections),
but its experimental value is a bit lower than the best fit SM value.


In all cases (except when $\sin\beta$ is close to $0.65$) the strongest bound on $M$ comes from asymmetries in $Z$ decays,
that mainly depend on the `effective $\sW^2$' that parameterizes the leptonic couplings of the $Z$.
An improvement in its measure  (or a shift in its central value) would thus affect our results.
An higher central value (like the one measured at LEP) would give stronger constraints,
while a lower value (like the one measured at SLD) could even indicate the presence of
a signal, if the error will be reduced by a factor $2\div 3$.
On the contrary no significant improvement of the bound on the compactification scale $M$ will result from
an improved measurement of the $W$ mass:
even with a $\pm15\MeV$ error the measure of $M_W$ will continue to give a subdominant bound.
Atomic parity violation gives negligible bounds on extra dimensions, and receives negligible corrections.

Comparable bounds are present in more general models, for example in presence of a single extra dimension
with the substructure proposed in~\cite{HS} to suppress proton decay.
With more than one extra dimension KK modes with mass close to the string scale give the dominant effect,
so that the details of the string `model' affect EWPO.
The larger multiplicity of KK modes probably implies stronger bounds on the compactification radii.

These LEP bounds make extremely unlikely that KK effects can be observed at Tevatron.
On the contrary LHC with high luminosity ($100\hbox{fb}^{-1}$)
can see KK effects  up to $M\sim (6\div 7)\TeV$~\cite{colliders5d}.
If a signal will be found, it could be difficult to distinguish directly KK modes from a compositness excitation of the $W$.
For example the effects of the next KK levels ($n>1$) could be too small to be seen.

\paragraph{Note added}
The same kind of analysis presented in section 3 has been performed in a recent paper~\cite{RW}.
Our conclusions and our bounds on $M$ agree with their results.

\paragraph{Acknowledgments}
We are grateful to Riccardo Barbieri and Riccardo Rattazzi for many useful discussions.

\appendix

\section{NRO operators}
Here we briefly recall few technical details, discussed in a more complete way  in~\cite{NRO},
where notations are precisely defined.

The following 10 operators are a minimal set of  gauge invariant, flavour symmetric,
CP-even operators of dimension 6 that can affect the EWPO:
\begin{eqnarray*}
\Ord_{WB}&=&(H^\dagger \tau^a H) W^a_{\mu\nu} B_{\mu\nu} \\
\Ord_{H}&=&|H^\dagger D_\mu H|^2 \\
\Ord_{LL}&=&{\textstyle \frac{1}{2}}(\bar{L}\gamma_\mu \tau^a L)^2 \\
\Ord_{HL}' &=&i(H^\dagger D_\mu \tau^a H)(\bar{L}\gamma_\mu \tau^a L) +\hc\\
\Ord_{HQ}' &=&i(H^\dagger D_\mu \tau^a H)(\bar{Q}\gamma_\mu \tau^a Q) +\hc\\
\Ord_{HL} &=&i (H^\dagger D_\mu H)(\bar{L}\gamma_\mu L) +\hc\\
\Ord_{HQ} &=&i (H^\dagger D_\mu H)(\bar{Q}\gamma_\mu Q)+\hc \\
\Ord_{HE} &=& i (H^\dagger D_\mu H)(\bar{E}\gamma_\mu E)+\hc \\
\Ord_{HU} &=& i (H^\dagger D_\mu H)(\bar{U}\gamma_\mu U)+\hc\\
\Ord_{HD} &=& i (H^\dagger D_\mu H)(\bar{D}\gamma_\mu D) +\hc
\end{eqnarray*}
This set is minimal in the sense that any other operator that contributes to the EWPO of table~1
can be written as a combination of them, up to operators that give null contribution.
For our purposes the most general effective Lagrangian with dimension six operators can thus be written as
$${\cal L}_{\rm NRO}=\sum_{i=1}^{10} c_i {\cal O}_i.$$
The coefficients $c_i$ appropriate for our analysis are given in eq.\eq{LO}.
The contributions from the single operators to the
form factors $\delta e_i$, $\delta G_{\rm VB}$, $\delta g_{Vf}$ and $\delta g_{Af}$ (precisely defined, e.g., in~\cite{BCF})
are listed in table~2 of~\cite{NRO}.
The form factors affect the EWPO in an obvious way;
since the computation is however not immediate, the
explicit expressions given in~\cite{NRO} could be useful.
Fitting the $\epsilon_i$ (or the $S, T, U$) parameters would be much simpler;
however in presence of `non universal' corrections it is not a correct approximation and
a more cumbersome fit of the the EWPO is necessary.

The two operators
$$
\Ord_{WW}=\frac{1}{2}(D_\rho W^a_{\mu\nu})^2 ,\qquad
\Ord_{BB}=\frac{1}{2}(\partial_\rho B_{\mu\nu})^2
$$
can be written as a combination of the ten operators listed above,  plus operators
that do not affect EWPO.
They are however interesting for our analysis (see eq.s\eq{LO} and\eq{WWeBB}).
If they are present in the effective Lagrangian with coefficients $c_{WW}$ and $c_{BB}$,
the form factors receive the  following corrections
\begin{eqnarray*}
\delta e_2&=&c_{WW} g_2^2v^2 \tan^2\theta_{\rm W},\\
\delta e_4&=& +g_2^2v^2(c_{BB}+c_{WW}\tan^2\theta_{\rm W}),\\
\delta e_5&=&-g_2^2v^2(c_{WW}+c_{BB}\tan^2\theta_{\rm W})
\end{eqnarray*}
Since few of these form factors were zero in~\cite{NRO},
eq.~(2) of~\cite{NRO} must be generalized as~\cite{ABC}
\begin{eqnarray*}
\delta \epsilon_1 &=& \delta e_1 - \delta e_5-\delta G_{\rm VB}\\
\delta \epsilon_2 &=& \delta e_2 - s^2_{\rm W} \delta e_4- c^2_{\rm W}\delta e_5-\delta G_{\rm VB}\\
\delta \epsilon_3 &=& \delta e_3 + c^2_{\rm W} \delta e_4 -c^2_{\rm W}\delta e_5.
\end{eqnarray*}


\frenchspacing
\small\footnotesize
\begin{thebibliography}{nn}

\bibitem{d}
Compactified extra dimensions were proposed in
\art{T. Kaluza}{Preuss. Akad. Wiss.}{}{966}{1921};
\art{O. Klein}{Z. Phys.}{37}{895}{1926}.
For more recent works see:
\art{I. Antoniadis}{\PL}{B246}{377}{1990};
\art{E. Witten}{\NP}{B471}{135}{1996};
\art{J.D. Lykken}{\PR}{D54}{3693}{1996}.
The fact that large `gravitational' dimensions, not excluded by any experiment,
allow TeV-scale strings  has been observed in
\art[hep-ph/9803315]{N. Arkani-Hamed, S. Dimopoulos and G. Dvali}{\PL}{B429}{263}{1998};
\art[hep-ph/9804398]{I. Antoniadis, N. Arkani-Hamed, S. Dimopoulos and G. Dvali}{\PL}{B436}{263}{1998};
\art[hep-ph/9807344]{N. Arkani-Hamed, S. Dimopoulos and G. Dvali}{\PR}{D59}{086004}{1999}.
Unification of couplings in presence of extra dimensions has been studied in~\cite{DDG}.


\bibitem{NRO} \hepart[hep-ph/9905281]{R. Barbieri and A. Strumia}.



\bibitem{ColliderGravitonico}
Computable graviton effects have been studied in
\art[hep-ph/9811291]{G. Giudice, R. Rattazzi and J. D. Wells}{\NP}{B544}{3}{1999};
\hepart[hep-ph/9811337]{E.A. Mirabelli, M. Perelstein and M.E. Peskin};
\hepart[hep-ph/9811356]{J.L. Hewett};
The angular distribution of the effect at $e^+e^-$ colliders
has been studied in
\hepart[hep-ph/9902263]{K. Agashe and N.G. Deshpande}.

\bibitem{supernova} 
See the last ref. in~\cite{d} for an estimation.
\hepart[hep-ph/9904267]{L. J. Hall and D. Smith};
\hepart[hep-ph/9905474]{V. Barger, T. Han, C. Kao and R.J. Zhang}.



\bibitem{DDG}
Unification in extra dimensions has been studied in
\art{K.R. Dienes, E. Dudas and T. Gherghetta}{\PL}{B43}{55}{1998}.


\bibitem{NY}
\hepart[hep-ph/9902323]{P. Nath and M. Yamaguchi}.
Stronger bounds on the compactification scale
are quoted in the subsequent analyses of precision data, ref.s~\cite{MP} and~\cite{Marciano}.
Bounds from EWPO have also been discussed in
\hepart[hep-th/9903019]{T. Banks, M. Dine and A.E. Nelson}.

\bibitem{MP} \hepart[hep-ph/9902467]{M. Masip and A. Pomarol}.


\bibitem{Marciano} 
\hepart[hep-ph/9903451]{W. J. Marciano}.





\bibitem{M5SM}
\art[hep-ph/9806263]{A. Pomarol and M. Quir\'os}{\PL}{B438}{255}{1998}.
See also \cite{MP}.

\bibitem{mh<250} LEP working group.
Avaible  at the www address
{\tt www.cern.ch/LEPEWWG/plots/winter99}.
The full set of data involves measurements at LEP, SLC and Tevatron.

\bibitem{diLella} L. di Lella, experimental summary of the 34th recontres de Moriond
on electroweak interactions and unified theories.
Transparencies available at the www address {\tt moriond.in2p3.fr/EW/transparencies.}

\bibitem{Csexp}
\hepart[hep-ex/9903022]{S.C. Bennett and C.E. Wiemann}.


\bibitem{QWexp}
\art{C.S. Wood et al.}{Science}{275}{1759}{1997}.


\bibitem{QWfit}
\art[hep-ph/9905568]{R. Casalbuoni, S. De Curtis, D. Dominici, R. Gatto}{\PL}{B460}{135}{1999}. 

\bibitem{GW} \art{B. Grinstein and M.B. Wise}{\PL}{B265}{326}{1991}.

\bibitem{Carone} \hepart[hep-ph/9907362]{C.D. Carone}.

\bibitem{Casalbuoni}
\art[hep-ph/9805446]{R. Casalbuoni, S. De Curtis, D. Dominici, R. Gatto and M. Grazzini}{\PL}{B435}{396}{1998}.

\bibitem{HK} \hepart[hep-ph/9904236]{L. Hall and C. Kolda}.

\bibitem{dmhSM}
\art{M. Veltman}{Acta Phys. Pol.}{B12}{437}{1981}.

\bibitem{pdg}
Review of particle physics, European Phys. Journal {\bf C3} (1998) 1
(also available  at the www address www.pdg.lbl.gov).

\bibitem{ABC} 
for a review see
\art[hep-ph/9712368]{R. Barbieri, F. Caravaglios, G. Altarelli}{Int. J. Mod. Phys.}{A13}{1031}{1998}
and ref.s therein.

\bibitem{2h}
\art[hep-ph/9307201]{H.E. Haber and R. Hempfling}{\PR}{D48}{4280}{1993}.


\bibitem{STU} \art{M. Peskin and T. Takeuchi}{\PRL}{65}{194}{1990}
\art{}{\PR}{D46}{381}{1991}.

\bibitem{BCF} \art{R. Barbieri, F. Caravaglios, M. Frigeni}{\PL}{B279}{169}{1992}.


\bibitem{HS} \hepart[hep-ph/9903417]{N.Arkani-Hamed and M. Schmaltz}.


\bibitem{colliders5d}
\art{I. Antoniadis, K. Benakli and M. Quir\' os}{\PL}{B331}{313}{1994};
\hepart[hep-ph/9905311]{I. Antoniadis, K. Benakli and M. Quir\' os};
\hepart[hep-ph/9905415]{P. Nath, Y. Yamada and M. Yamaguchi}.
See also~\cite{RW}.


\bibitem{RW} \hepart[hep-ph/9906234]{T. G. Rizzo and J.D. Wells}.




\end{thebibliography}
\end{document}